\documentclass[conference]{IEEEtran}
\usepackage{cite}

\usepackage{hyperref}
\hypersetup{
    colorlinks=true,
    linkcolor=blue,
    filecolor=magenta,      
    urlcolor=cyan,
    pdftitle={Overleaf Example},
    pdfpagemode=FullScreen,
    }

\usepackage[cmex10]{amsmath}
\usepackage{amssymb,amsfonts}
\usepackage{mathdots}
\usepackage{cases}
\usepackage{eqparbox}

\usepackage[pdftex]{graphicx}

\usepackage{fixltx2e}

\newcommand{\diag}{\mathop{\mathrm{diag}}}

\begin{document}
\title{A low multiplicative complexity fast recursive DCT-2 algorithm}

\author{\IEEEauthorblockN{Maxim Vashkevich}
\IEEEauthorblockA{Computer Engineering Department\\
Belarusian State University \\of Informatics and Radioelectronics\\
Minsk, Belarus, 220013\\
Email: vashkevich@bsuir.by}
\and
\IEEEauthorblockN{Alexander Petrovsky}
\IEEEauthorblockA{Computer Engineering Department\\
Belarusian State University \\of Informatics and Radioelectronics\\
Minsk, Belarus, 220013\\
Email: palex@bsuir.by}}


\maketitle

\begin{abstract}
A fast Discrete Cosine Transform (DCT) algorithm is introduced that can be of particular interest in image processing. The main features  of the algorithm are regularity of the graph and very low arithmetic complexity. The 16-point version of the algorithm requires only 32 multiplications and 81 additions. The computational core of the algorithm consists of only 17 nontrivial multiplications, the rest 15 are scaling factors that can be compensated in the post-processing. The derivation of the algorithm is based on the algebraic signal processing theory (ASP). MATLAB implementation of the algorithm can be found in the public repository \url{https://github.com/Mak-Sim/Fast_recursive_DCT}
\end{abstract}

%
\IEEEpeerreviewmaketitle

\section{Introduction}
The Discrete Cosine Transform (DCT) has found many applications in image processing, data compression and other fields due to its decorrelation property~\cite{Rao1990}. Despite the fact that a number of fast DCT algorithms has been proposed already, designing new efficient schemes is still of great interest~\cite{Parf2006, Liang01, Vash09}. Majority of proposed fast DCT algorithms have been obtained using graph transformation, equivalence relation or sophisticated manipulation of the transform coefficients. Recently an algebraic approach to derivation of fast DCT has been presented~\cite{Pusch3}. The approach uses {\it polynomial algebra} associated with DCT to obtain fast algorithms. Subsequently this theory has been called {\it algebraic signal processing theory} (ASP)~\cite{Pusch8d}. The theory provides consistent algebraic interpretation of fast DCT algorithms.

The paper presents derivation of fast DCT-2 $n$-point algorithm  ($n$ is a power of two) based on ASP. The algorithm is recursive and has a regular graph. Another feature of the algorithm is very low arithmetic complexity: 16-point DCT requires only 32 multiplications and 81 additions (that is only one multiplication greater than~\cite{Loeff89}), but the computational core of algorithm contains only 17 multiplication while other 15 are scaling factors that can be compensated in the post-processing. Because of the mentioned properties the algorithm is a very attractive choice for hardware DCT implementations.

\section{Algebraic approach to DCT}
In this section the fundamentals of algebraic signal processing theory~\cite{Pusch8d} are considered that are used further for derivation of the fast DCT-2 algorithm.

\subsection{Background: polynomial algebras}
A {\it polynomial algebra} is a vector space over the field $\mathbb{F}$ denoted as
\begin{equation}
\mathcal{A}_{\mathbb{F}} = \mathbb{F}[x]/p(x). \label{eq:pol_alg}
\end{equation}
The elements of algebra is the set of all polynomials in $x$ over $\mathbb{F}$ of degree smaller than $\deg(p)=n$. $\mathcal{A}_{\mathbb{F}}$ is equipped with the operations of usual polynomial addition and multiplication modulo the polynomial $p(x)$.

Using the Chinese remainder theorem (CRT) a polynomial algebra~(\ref{eq:pol_alg}) can be decomposed into a direct sum of one-dimensional subalgebras 
\begin{equation}
\mathcal{F} \; \colon \;\; \mathbb{F}[x]/p(x)\rightarrow \bigoplus_{0\leq k < n} \mathbb{F}_e[x]/(x-\alpha_k),
\label{eq:CRT}
\end{equation} 
provided that zeros $\alpha=(\alpha_0,\ldots,\alpha_{n-1})$ of $p(x)$ are pairwise distinct and $\alpha_k\in\mathbb{F}$. The mapping $\mathcal{F}$ is represented in matrix form
\begin{equation}
\mathcal{F} = \mathcal{P}_{b,\alpha}=[p_\ell(\alpha_k)]_{0\leq k,\ell <n},
\label{eq:pol_trans}
\end{equation}
if a basis $b=(p_0,\ldots,p_{n-1})$ is set in $\mathbb{F}[x]/p(x)$ and unit bases $(x^0)=(1)$ is chosen in each $\mathbb{F}[x]/(x-\alpha_k)$. $\mathcal{P}_{b,\alpha}$ is referred to as {\it polynomial transform} for $\mathcal{A}_{\mathbb{F}}$ with basis~$b$~\cite{Pusch8d}. A {\it scaled polynomial transform} is obtained for a different basis $\beta_k$ in each $\mathbb{F}[x]/(x-\alpha_k)$:
\begin{equation}
\mathcal{F} =\diag(1/\beta_1,\ldots,1/\beta_{n-1}) \cdot \mathcal{P}_{b,\alpha}.
\label{eq:sc_pol_trans}
\end{equation}

\subsection{Derivation of fast transform algorithms in ASP}
In ASP transforms is represented as matrix-vector products
\begin{equation*}
\mathbf{y} = T \mathbf{x}, \text{ where } T=[t_{k,\ell}]_{0\leq k,\ell<n}.
\end{equation*}
The fast transform algorithm is viewed as factorization of $T$ into a product of sparse structured matrices. 
This approach has advantages from an algorithmic point of view. It reveals the algorithm structure and simplifies manipulation with it to derive a new variants.

In the paper the following basic matrices are used:
\begin{equation*}
I_n = \begin{bmatrix}
1	&			&		\\
	&	\ddots	& 		\\
	&			&	1
\end{bmatrix} \quad
J_n = \begin{bmatrix}
	&			&	1	\\
	&	\iddots	& 		\\
1	&			&	
\end{bmatrix}.
\end{equation*}
Permutation matrices that has exactly one entry 1 in row $i$ at position $f(i)$ and each column and 0 elsewhere is defined as:
\begin{equation*}
P\colon i\mapsto f(i),\quad 0\leq i < n.
\end{equation*}
One important is the $n \times n$ stride permutation matrix defined for $m|n$ as
\begin{equation*}
L_m^n \colon i_2\frac{n}{m}+i_1 \mapsto i_1 m + i_2
\end{equation*}
for $0 \leq i_1< \frac{n}{m}$, $0 \leq i_2 < m$.

ASP states that every DCT corresponds to some polynomial algebra $\mathbb{F}[x]/p(x)$ with basis $b$. In this case DCT is given by the CRT~(\ref{eq:CRT}) and its matrix takes the form of polynomial transform~(\ref{eq:pol_trans}) or a scaled polynomial transform~(\ref{eq:sc_pol_trans}). From~(\ref{eq:CRT}) it can be seen that $\mathcal{F}$ decomposes $\mathbb{F}[x]/p(x)$ into one-dimensional polynomial algebras. Fast algorithm is obtained by complying this decomposition {\it in step} using an intermediate subalgebras.

One possible way to perform decomposition of $\mathbb{F}[x]/p(x)$ in step is to use factorization $p(x)=q(x)\cdot r(x)$. If $\deg(q)=k$ and $\deg(r)=m$ then
\begin{eqnarray}
&& \mathbb{F}[x]/p(x) \notag \\
& \rightarrow &  \mathbb{F}[x]/q(x)\oplus \mathbb{F}[x]/r(x) \label{eq6}\\
& \rightarrow & \bigoplus\limits_{0\leq i<k}\mathbb{F}[x]/(x-\beta_i) \oplus \bigoplus\limits_{0\leq j<m} \mathbb{F}[x]/(x-\gamma_j) \label{eq7}\\
& \rightarrow & \bigoplus\limits_{0\leq i<n}\mathbb{F}[x]/(x-\alpha_i) \label{eq8}
\end{eqnarray}
where $\beta_i$ and $\gamma_j$ are the zeros of $q(x)$ and $r(x)$ correspondingly. If $c$ and $d$ are the bases of $\mathbb{F}[x]/q(x)$ and $\mathbb{F}[x]/r(x)$, respectively, then (\ref{eq6})-(\ref{eq8}) are expressed in the following matrix form~\cite{Pusch8d}:
\begin{equation}\label{th1}
\mathcal{P}_{b,\alpha} = P(\mathcal{P}_{c,\beta}\oplus \mathcal{P}_{d,\gamma})B,
\end{equation}
where $A\oplus B=[\begin{smallmatrix} A & \\ & B\end{smallmatrix}]$ denotes the direct sum of matrices. Step~(\ref{eq7}) uses the CRT to decompose $\mathbb{F}[x]/q(x)$ and $\mathbb{F}[x]/r(x)$. This step corresponds to the direct sum of matrices $\mathcal{P}_{c,\beta}$ and $\mathcal{P}_{d,\gamma}$. Finally permutation matrix $P$ maps the concatenation $(\beta,\gamma)$ to the ordered list of zeros $\alpha$ in (\ref{eq8}). Given that $B$ is sparse~(\ref{th1}) leads to a fast algorithm.

\subsection{Polynomial algebras for $\mathrm{DCT}$-$2$ and $\mathrm{DCT}$-$4$}%
This subsection introduces polynomial algebras which is connected with $\mathrm{DCT}\text{-}4$ and $\mathrm{DCT}\text{-}2$.
Let us first consider the polynomial algebra associated with $\mathrm{DCT}\text{-}4_n$
\begin{equation}
\mathcal{A}_\mathbb{F} = \mathbb{F}[x]/2T_n(x),\quad b=(V_0,\ldots,V_{n-1}), \label{eq:dct4_pol_albr}
\end{equation}
where 
$T$ and $V$ are Chebyshev polynomials of the first and third kind, respectively. This Chebyshev polynomials have the following closed form expressions ($\cos\theta=x$)
\begin{equation*} \textstyle
T_n(x) = \cos(n\theta),\,\,\, V_n(x) = \frac{\cos(n+\frac12)\theta}{\cos\frac12\theta}.
\end{equation*}
$\alpha_k=\cos(k+\frac12)\frac{\pi}{n}$, $0\leq k<n$ are zeros of $2T_n(x)$. in accordance with~(\ref{eq:pol_trans}) polynomial transform for algebra (\ref{eq:dct4_pol_albr})  is defined as
\begin{equation} \label{eq4}
\mathcal{P}_{\alpha,b} = [ V_\ell(\alpha_k)]_{0\leq k,\ell<n} = \left[  \frac{\textstyle \cos(k+\frac{1}{2})(\ell+\frac{1}{2})\frac{\pi}{n}} {\textstyle\cos(k+\frac{1}{2})\frac{\pi}{2n}} \right].
\end{equation}
In order to get the matrix of $\mathrm{DCT}$-$4_n$ (\ref{eq4}) is multiplied  from the left by scaling diagonal matrix
\begin{equation*} \label{eq:sc_mtx}\textstyle
D_n^{(C4)} = \diag\nolimits_{0\leq k< n} \left({\cos(k+\frac12)\frac{\pi}{2n}}\right)
\end{equation*}
that yields
\begin{equation} \label{eq5}\textstyle
\mathrm{DCT}\text{-}4_n =\left[\cos(k+\frac12)(\ell+\frac12)\frac{\pi}{n} \right]_{0\leq k,\ell<n}.
\end{equation}
Eq. (\ref{eq4})--(\ref{eq5}) show that $\mathrm{DCT}$-$4$ is a scaled polynomial transform of the form (\ref{eq:sc_pol_trans}) for the specified polynomial algebra~(\ref{eq:dct4_pol_albr}). 

$\mathrm{DCT}\text{-}2_n$ is arisen from polynomial algebra 
\begin{equation}
\mathcal{A}_\mathbb{F} = \mathbb{F}[x]/(x-1)U_{n-1}(x),\quad b=(V_0,\ldots,V_{n-1}), \label{eq:dct2_pol_albr}
\end{equation}
where $U$ is Chebyshev polynomial of the second kind that can be written as ($\cos\theta=x$):
\begin{equation*} \textstyle
U_n(x) = \frac{\sin(n+1)\theta}{\sin\theta}.
\end{equation*}
Since zeros of $U_n(x)$ is given by $\alpha_k=\cos\frac{(k+1)\pi}{n+1}$, $0\leq k<n$ polynomial transform for~(\ref{eq:dct2_pol_albr}) takes the form
\begin{equation} \label{eq:dct2_pol_trans}
\mathcal{P}_{\alpha,b} = [ V_\ell(\alpha_k)]_{0\leq k,\ell<n} = \left[  \frac{\textstyle \cos k(\ell+\frac{1}{2})\frac{\pi}{n}} {\textstyle\cos\frac{k\pi}{2n}} \right].
\end{equation}
To obtain $\mathrm{DCT}\text{-}2$ matrix (\ref{eq:dct2_pol_trans}) need to be multiplied from the left by the scaling diagonal
\begin{equation*} \label{eq:sc_mtx_c2}\textstyle
D_n^{(C2)} = \diag\nolimits_{0\leq k< n} \left({\cos\frac{k\pi}{2n}}\right).
\end{equation*}
Polynomial transform corresponding to discrete trigonometric transform ($\mathrm{DTT}$) is denoted as $\overline{\mathrm{DTT}}$, for instance $\overline{\mathrm{DCT}\text{-}4}_n$ stands for the matrix in (\ref{eq4}). 

In what follows we need skew $\mathrm{DCT}\text{-}4(r)$. In~\cite{Pusch8d} this transform was introduced since it appears to be important building blocks of Cooley-Tukey type of algorithms for $\mathrm{DCT}$. Skew $\mathrm{DCT}\text{-}4(r)$ associates with polynomial algebra 
\[ \mathcal{A}_{\mathbb{F}}=\mathbb{F}[x]/(2T_n(x)-2\cos r\pi)\] 
with the basis $b=(V_0,\ldots,V_{n-1})$, where $0<r<1$. The conventional $\mathrm{DCT}\text{-}4_n$ is the special case of skew $\mathrm{DCT}\text{-}4_n(r)$ for $r=1/2$.

\section{Derivation of fast $\mathrm{DCT}\text{-}2_{2^k}$ algorithm}%
In this section the procedure of algebraic derivation of fast $\mathrm{DCT}\text{-}2_{2^k}$ algorithm is given in detail. According to~(\ref{eq:dct2_pol_albr}) the polynomial algebra corresponding to $\mathrm{DCT}\text{-}2_{2^k}$ is given by
\begin{equation*}
\mathcal{A}_\mathbb{F} = \mathbb{F}[x]/(x-1)U_{2^k-1}(x),\quad b=(V_0,\ldots,V_{2^k-1}).
\end{equation*}
Important issue is to choose the base field $\mathbb{F}$. Since Chebyshev polynomials $V$ and $U$ which is included in definition (\ref{eq:dct2_pol_albr}) have integer coefficients (for example $V_2(x)=4x^2-2x-1$), the base field $\mathbb{F}$ is set to the field of rational numbers $\mathbb{Q}$. The filed is extended during factorization of polynomial $U_{2^k-1}(x)$, since the polynomial is not factored over $\mathbb{Q}$.

It is well known \cite{Rao1990} that fast $\mathrm{DCT}\text{-}2_{2n}$ algorithm can be reduced to fast $\mathrm{DCT}\text{-}2_{n}$ and $\mathrm{DCT}\text{-}4_{n}$ algorithms. Using factorization for the Chebyshev polynomial of the second kind 
\begin{equation*}
U_{2n-1}(x) = U_{n-1}(x)\cdot 2T_n(x),
\end{equation*}
the algebra $\mathbb{Q}[x]/(x-1)U_{2n-1}(x)$ with basis $b=(V_0,\ldots $ $V_{2n-1})$ can be decomposed as
\begin{eqnarray}
&&\mathbb{Q}[x]/(x-1)U_{2n-1}(x) \notag \\ 
&\rightarrow&\mathbb{Q}[x]/(x-1)U_{n-1}(x)\oplus \mathbb{Q}[x]/2T_{n}(x), \label{eq:dct2_2n_decomp}
\end{eqnarray}
that according to (\ref{eq6})--(\ref{eq8}) leads to the following fast  algorithm~\cite{Pusch8d}
\begin{equation}\label{eq:fast_dct2}
\overline{\mathrm{DCT}\text{-}2}_{2n} = L_n^{2n}(\overline{\mathrm{DCT}\text{-}2}_{n}\oplus \overline{\mathrm{DCT}\text{-}4}_{n})B_{2n},
\end{equation}
where $L_n^{2n}$ is the stride permutation matrix and $B_{2n}$ is change of basis matrix. $B_{2n}$ maps basis $b$ to the concatenation $(c,d)$, where $c=d=(V_0,\ldots V_{n-1})$ are the basis for subalgebras in the right-hand side of~(\ref{eq:dct2_2n_decomp}). The first $n$ columns of $B_{2n}$ are 
\begin{equation*}
B_{2n}=\begin{bmatrix}
I_n	&	*	\\
I_n	&	*
\end{bmatrix},
\end{equation*}
since the elements $V_\ell \in b$ for $0\leq \ell < n$ are already contained in $c$ and $d$. The rest entries are determined by the following expressions
\begin{eqnarray}
V_{n+\ell} \equiv&    V_{n-\ell-1}	 &\mod{(x-1)U_{n}}  \label{eq:new_dct2_basis:a}\\ 
V_{n+\ell} \equiv& 	-V_{n-\ell-1}	 &\mod{2T_{n}},  \label{eq:new_dct2_basis:b}
\end{eqnarray}
which yields
\begin{equation*}
B_{2n}=\begin{bmatrix}
I_n	&	 J_n	\\
I_n	&	-J_n
\end{bmatrix}.
\end{equation*}
(\ref{eq:new_dct2_basis:a})--(\ref{eq:new_dct2_basis:b}) can be induced using the following relation $2T_n = V_n + V_{n-1}$, $(x-1)U_{n-1} = V_n - V_{n-1}$ and $V_n = 2xV_{n-1}-V_{n-2}$. Note that decomposition~(\ref{eq:dct2_2n_decomp}) does not require extension of based field $\mathbb{Q}$. This leads to multiplication-free change of basis matrix $B_{2n}$. 

When the size of $\mathrm{DCT}\text{-}2$ is power of two (\ref{eq:fast_dct2}) can be applied recursively to obtain fast algorithm. Thus, the problem of derivation of fast $\mathrm{DCT}\text{-}2_{2^k}$ algorithm reduces to derivation of fast $\mathrm{DCT}\text{-}4_{2^{k-1}}$ algorithm. From the ASP point of view the question is how to factor polynomial $2T_n$ (when $n$ is power of 2) in step. We propose to use the following general recursive formula
\begin{multline}
\textstyle 2T_{2n}(x) -2\cos r\pi = \mathstrut\left(2T_{n}(x)-2\cos\frac{r\pi}{2}\right) \\ \textstyle 
\times\left(2T_{n}(x)-2\cos \pi(1-\frac{r}{2})\right), \label{eq:gen_fact}
\end{multline} 
that can be proved using the closed form of $T_{2n}$, parameter $r\in(0,\;1)$. The special case of~(\ref{eq:gen_fact}) for $r=1/2$ specify factorization of $2T_{2n}$. Using (\ref{eq:gen_fact}) polynomial algebra related to $\mathrm{DCT}\text{-}4_{2n}(r)$ is decomposed as\footnote{Here $\mathbb{Q}_{\cos r\pi}$ is used as a short notation for field extension $\mathbb{Q}[\cos r\pi]$.} 
\begin{eqnarray}
&& \mathbb{Q}_{\cos r\pi}[x]/(2T_{2n}(x)-2\cos r\pi) \notag \\
& \rightarrow &  \textstyle \mathbb{Q}_{\cos\frac{r\pi}{2}}[x]/(2T_{n}(x)-2\cos\frac{r\pi}{2}) \oplus \notag\\
&& \textstyle \mathbb{Q}_{\cos\frac{r\pi}{2}}[x]/(2T_{n}(x)-2\cos \pi(1-\frac{r}{2})). \label{eq:dct4_2n_decomp}
\end{eqnarray}
The decomposition leads to the following fast algorithm
\begin{equation}\label{my_alg}
\begin{split}
\textstyle 
\overline{\mathrm{DCT}\text{-}4}_{2n}(r) =  P\cdot(& \textstyle\overline{\mathrm{DCT}\text{-}4}_{n}(\frac{r}{2})  \\ \oplus&\textstyle\overline{\mathrm{DCT}\text{-}4}_{n}(1-\frac{r}{2}) )\cdot B_{2n}^{(C4)}(r),
\end{split}
\end{equation}
where $P$ is a permutation matrix of the form
\begin{equation*}
P= \left[
\begin{smallmatrix} 
	1	&		& 			&	&			&			&		\\
		&		&			& 	&	I_2	& 			&		\\
		&	I_2&			&	&	 		&			&		\\
		&		&			&	&	 		&\ddots	&		\\
		& 		&\ddots	&	&			&			&		\\
		&		&			&	&			&			&	I_2\\
		&		&			& 1&			&			&	
\end{smallmatrix}\right],
\end{equation*}
and $B_{2n}^{(C4)}(r)$ is the change of basis matrix
\begin{equation}
\begin{split}
B_{2n}^{(C4)}(r)=&
\begin{bmatrix}
I_m	&	(2\cos\frac{r\pi}{2}I_m-J_m)	\\
I_m	&	(-2\cos\frac{r\pi}{2}I_m-J_m)
\end{bmatrix} \\
=&\begin{bmatrix}
I_m	&	I_m	\\
I_m	&	-I_m
\end{bmatrix}\cdot
\begin{bmatrix}
I_m	&	-J_m	\\
		&	2\cos\frac{r\pi}{2}I_m
\end{bmatrix},
\end{split}\label{eq:BB_matrix}
\end{equation}
which is determined by
\begin{equation*}
\begin{split}\textstyle
V_{n+\ell} \equiv\textstyle -  V_{n-\ell-1} + & \textstyle 2\cos\frac{r\pi}{2}V_\ell   
		 \textstyle \mod{2T_{n} - 2\cos\frac{r\pi}{2}}  
\\ \textstyle
V_{n+\ell} \equiv\textstyle -  V_{n-\ell-1} - & \textstyle 2\cos\frac{r\pi}{2}V_\ell  
		 \textstyle\mod{2T_{n} - 2\cos \pi(1-\frac{r}{2})}.
\end{split}
\end{equation*}

Decomposition~(\ref{eq:dct4_2n_decomp}) requires extension of the based field $\mathbb{Q}_{\cos r\pi}$ to $\mathbb{Q}_{\cos\frac{r\pi}{2}}$. New elements of the field appears in matrix $B_{2n}^{(C4)}(r)$. 

Joint use of factorizations (\ref{eq:fast_dct2}) and (\ref{my_alg}) leads to the new fast $\mathrm{DCT}\text{-}2_{2^k}$ recursive algorithm. The basic operation of the algorithm is multiplication by the matrix $B_{2n}^{(C4)}(r)$. All nontrivial multiplication concentrate in it that is very similar to butterfly operation in FFT algorithm.

\begin{figure*}[!t]
\centering\includegraphics[width=160mm]{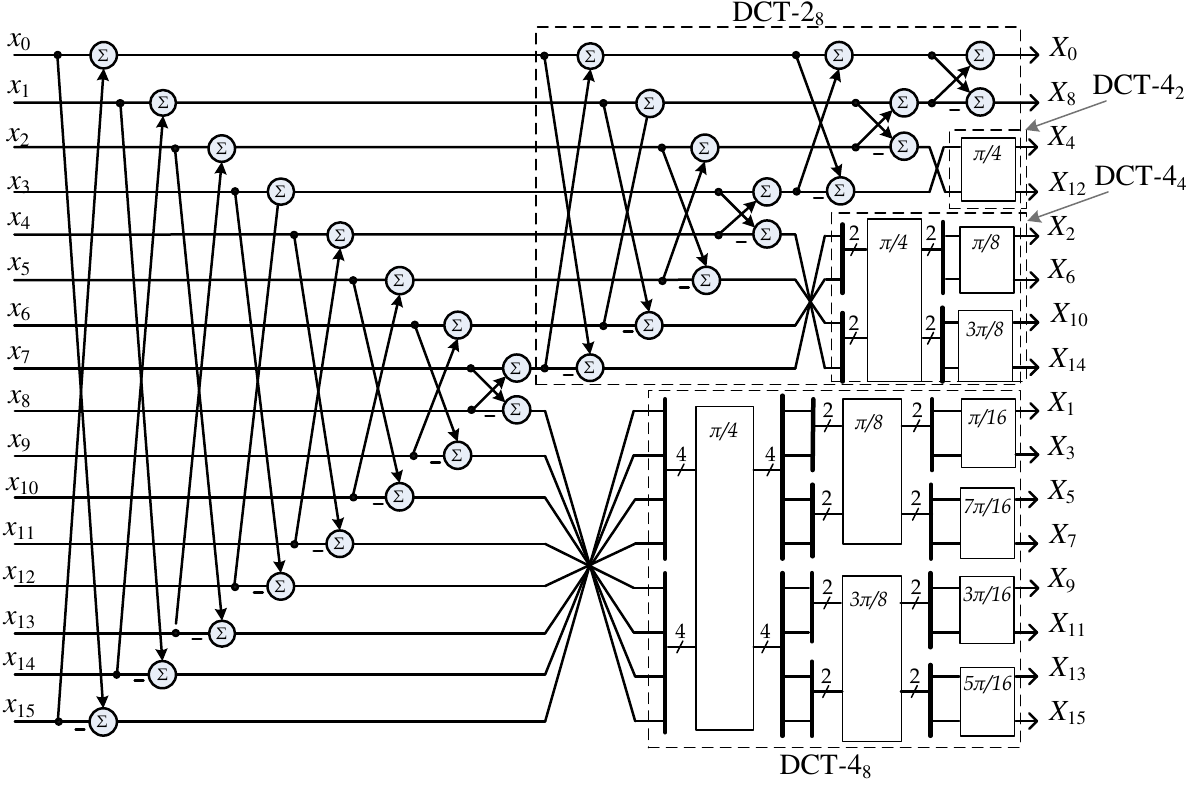}
\caption{16-point DCT algorithm}\label{fig:1}
\end{figure*}

\begin{figure}[tb]
\centering\includegraphics[width=72mm]{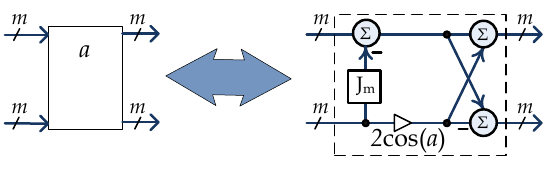}
\caption{Building block of the fast DCT algorithm}\label{fig:2}
\end{figure}
\section{Fast $\mathrm{DCT}\text{-}2_{16}$ algorithm} 				%
In this section the proposed approach is applied to derivation of fast $\mathrm{DCT}\text{-}2_{16}$ algorithm. At first the transform expressed as a product 
\begin{equation}
\mathrm{DCT}\text{-}2_{16} = D_{16}^{(C2)} \cdot \overline{\mathrm{DCT}\text{-}2}_{16}.
\end{equation}
Then factorization (\ref{eq:fast_dct2}) and (\ref{my_alg}) is applied recursively to obtain fast transform algorithm. Flow graph of this algorithm is shown in Fig. \ref{fig:1} (for simplicity scaling of the output is omitted). Fig. \ref{fig:2} explains the basic building block (BB) of the algorithm that performs the multiplication by matrix~(\ref{eq:BB_matrix}). All operations inside the BB are implemented on the input $m$ components vectors.  Evaluation of one BB requires $3m$ addition and $m$ multiplication. 

The presented 16-point $\mathrm{DCT}\text{-}2$ algorithm uses 32 multiplication and 81 addition~\footnote{You can find MATLAB implementation of the algorithm in the public repository \url{https://github.com/Mak-Sim/Fast_recursive_DCT}}. However only 17 multiplication constitute the core of algorithm while other 15 is scaling factors that can be compensate in the post-processing. Also the Fig.~\ref{fig:1} shows that algorithm include computation of 8-point $\mathrm{DCT}\text{-}2$ that requires only 5 multiplication. It is the same result as in~\cite{Arai88}. In fact, proposed algorithm can be considered as generalization of Arai's DCT algorithm since the resulting computational scheme has very low multiplicative complexity and scaling outputs.

\section{Conclusion}

A fast $2^k$-point algorithm of DCT-2 based on ASP is presented. The key features of the algorithm are regularity of the graph ($\mathrm{DCT}\text{-}2_{n/2}$ available inside of a $\mathrm{DCT}\text{-}2_n$) and very low arithmetic complexity (computational core of the $\mathrm{DCT}\text{-}2_{2^k}$ algorithm contains only $\sum_{p=1}^{k-1}2^pp$ multiplications). Regular graph of proposed algorithm is well suited for development of new parallel-pipeline architecture of DCT processor. Also, the algorithm extends existing space of alternative fast algorithms of the DCT. It can be used by automatic code generation programs that search alternative implementations for the same transform to find the one that is best tuned to desired platform~\cite{Wanh1999,Pusch:05}.



\section*{Acknowledgment}
This work was supported by the Belarusian Fundamental Research Fund (F11MS-037).



\bibliographystyle{IEEEtran}
\bibliography{IEEEabrv,pcs_ref} 

\end{document}